\documentclass[conference]{IEEEtran}
\usepackage{cite}
\usepackage{amsmath,amssymb,amsfonts}
\usepackage{graphicx}
\usepackage{textcomp}
\usepackage{xcolor}
\usepackage{hyperref}
\usepackage{caption}
\usepackage[T1]{fontenc}
\usepackage{balance}
\usepackage{multirow}

\begin{document}

\title{On the impact of pull request decisions\\ on future contributions}

\author{\IEEEauthorblockN{Damien Legay}
	\IEEEauthorblockA{\textit{Software Engineering Lab} \\
		\textit{University of Mons}\\
		Mons, Belgium \\
		damien.legay@umons.ac.be}
	\and
	\IEEEauthorblockN{Alexandre Decan}
	\IEEEauthorblockA{\textit{Software Engineering Lab} \\
		\textit{University of Mons}\\
		Mons, Belgium \\
		alexandre.decan@umons.ac.be}
	\and
	\IEEEauthorblockN{Tom Mens}
	\IEEEauthorblockA{\textit{Software Engineering Lab} \\
		\textit{University of Mons}\\
		Mons, Belgium \\
		tom.mens@umons.ac.be}
}

\maketitle

\begin{abstract}
	The pull-based development process has become prevalent on platforms such as \textsf{GitHub} as a form of distributed software development.
	Potential contributors can create and submit a set of changes to a software project through pull requests. 
	These changes can be accepted, discussed or rejected by the maintainers of the software project, and can influence further contribution proposals.
	As such, it is important to examine the practices that encourage contributors to a project to submit pull requests. Specifically, we consider the impact of prior pull requests on the acceptance or rejection of subsequent pull requests.
	We also consider the potential effect of rejecting or ignoring pull requests on further contributions.   
	In this preliminary research, we study three large projects on \textsf{GitHub}, using pull request data obtained through the \textsf{GitHub} API, and we perform empirical analyses to investigate the above questions.
	Our results show that continued contribution to a project is correlated with higher pull request acceptance rates and that pull request rejections lead to fewer future contributions.
\end{abstract}


\section{Introduction}

The turn of the century saw the rise of version control systems (\textit{VCS}) to support large-scale software engineering projects. Centralised VCS (e.g. \textsf{CVS} and \textsf{Subversion}) allow developers to share a common repository. Decentralised ones (e.g., \textsf{Mercurial} and \textsf{git}) allow each developer to own a local copy of the repository containing the full change history. This enables collaborative (often geographically distributed) software development on an hitherto unmatched scale. It has given birth to extremely popular online hosting platforms such as \textsf{GitHub}, \textsf{BitBucket} and \textsf{Mozdev}, allowing thousands of people to remotely work together on the same projects.
These platforms provide additional features on top of their underlying VCS to further support distributed collaborative development. Examples of such features are issue tracking, code review, integrated discussions, team management, documentation \& wiki, continuous integration and integration with external tools. 

Today, \textsf{git} has become the most popular distributed VCS by a large margin\footnote{For anecdotal evidence, based on a 2016 survey with 881 votes, 87\% of responders identified \textsf{git} as their VCS of choice \url{https://rhodecode.com/insights/version-control-systems-2016}}. It will thereby be the focus of our current research.
\textsf{git} supports two types of development processes:
 the \textit{shared repository} approach, where all contributors are given write access to the central repository and can therefore contribute to the project directly; and the \textit{pull request} (PR) approach where only project \textit{integrators} are allowed to do so.

With the PR approach, external contributions are managed indirectly: would-be contributors create a fork of the repository and, once they have addressed an issue or lack in the project, they request for their modifications to be ``pulled'' to the repository by submitting a \textit{pull request}. The project integrators can decide to approve these PRs, which are then merged into the main project's codebase.

PRs are extremely valuable, as they represent a major part of the project's continued evolution and expansion. It is therefore important to incentivise people to create pull requests, thereby contributing to the project. Previous studies have attempted to identify the factors influencing whether and when a PR will be merged~\cite{Gousios2014ESP,Gousios2015,Gousios2016}.

We expand upon this work, by focusing on determining those patterns of PR-acceptance behaviour that are indicative of continued contribution. Our working hypothesis is that people contributing to a project repository through PRs may get demotivated (and hence stop contributing) if their submitted PRs get rejected too often, or if too many of them are left open without any decision to merge them. Evolutionary insights in such phenomena may help us to understand which of such factors tend to dissuade people to keep contributing to a given project.

To this extent, we quantitatively study the following research questions using techniques based on survival analysis:

\textit{RQ$_1$: How are PR acceptance and rejection rates influenced by previous PRs?}
As a contributor accrues familiarity with a project, he becomes more able to contribute effectively, which we expect to result in a lower PR rejection rate. Similarly, as integrators become more acquainted with a contributor, they may develop a favourable bias towards his PRs, further decreasing rejection rates.

\textit{RQ$_2$: To which extent does PR acceptance or rejection influence further contributions?}

When his PRs are rejected, a developer could become discouraged and stop, temporarily or permanently, contributing to the project as a result. 

\textit{RQ$_3$: To which extent do PRs left open influence further contributions?}
A PR is sometimes left open for a long period, neither rejected nor merged into the core project. We posit this may constitute a form of "soft" rejection, wherein the integrators want to avoid alienating
the contributor but do not want to merge the PR.
Seeing a large number of untreated PRs may send an implicit message to potential contributors that the project integrators are unwilling or unable to process the volume of contributions they receive, and, therefore, that their participation to the project would not be valued or useful.

To provide preliminary evidence for these questions, we carry out an empirical analysis on a large number of PRs in three large, popular and long-lived projects on \textsf{GitHub}. We focus on \textsf{GitHub} because it is undoubtedly one of the largest and most active online hosting services for \textsf{git} projects.

\section{Related Work}

Several researchers have studied aspects related to the PR-based software development process,  either qualitatively or quantitatively.

Gousios and Zaidman proposed a PR dataset~\cite{Gousios2014dataset} including 900 projects and 350,000 PRs extracted using GHTorrent.
Through a mixed-method analysis of 291 \textsf{GitHub} projects, Gousios et al.~\cite{Gousios2014ESP} established that the PR-based development approach is used as frequently as the shared repository approach on \textsf{GitHub}. They observed that most PRs are short, receive few comments and are processed quickly. They also found that most PR rejections are due to the distributed nature of the pull-based process (e.g., PRs that are already obsolete upon creation).

In a follow-up work~\cite{Gousios2016}, they interviewed 645 contributors to examine their work practices and identify the challenges they face. They found that while contributors tend to check if their intended contribution is already covered, they do not communicate their intended contributions. 
Interviewed contributors outlined that poor responsiveness on the part of integrators could be a barrier to attracting or retaining contributors. 
Contributors also stated that it is hard to accept rejection of their PRs, as rejected PRs could harm their reputation as developers. Conversely, it is hard for integrators to explain the reasons for rejecting PRs.
Rejecting a PR without alienating its contributor was already identified as a challenge of the PR-based model~\cite{Gousios2015}. In that paper, they evaluated PRs from an integrator's point of view by interviewing 749 project integrators in order to understand which criteria are used to determine the quality of a PR and how they prioritise the evaluation of contributions. 
They found that most integrators decide to merge PRs based on project's objectives, their quality as measured by compliance to the project guidelines, test coverage and passing continuous integration checks. 

Yu et al. \cite{Yu2015msr} studied the factors that contribute to latency in PR reviews, defining this latency as the ``time interval between pull request creation and closing date''. 
They found that PR latency is mainly affected by process-related factors such as whether a PR was assigned to a specific reviewer or not. They also found that continuous integration is a dominant factor in PR latency.

Rahman and Roy \cite{Rahman2014msr} categorised the technical issues discussed in PR comments and analysed information about projects and developers to obtain insights into PR acceptance or rejection. 
They discovered that the rate of PR rejection is highly correlated to the programming language used (e.g., Java PRs are more frequently rejected than PRs for the C programming language), the application domain of the project (e.g., the database application domain sees fewer merged PRs than the IDE domain), the maturity of a project (older projects accept fewer PRs) and the number of developers on the project. 

Tsay et al.~\cite{Tsay2014ICSE} explored both technical and social factors that contribute to acceptance of PRs.
They found that, although technical factors like the presence of tests in the PR and a small number of lines changed contribute to a higher probability of acceptance, social factors, such as whether the contributor follows the user that closes the PR, had stronger associations to PR acceptance than technical ones.

Terrel et al.~\cite{Terrell2016} established that PR acceptance is subject to a bias against women, when their gender is identifiable.

Rastogi et al.~\cite{Rastogi2018ESEM} built upon the factors identified in \cite{Gousios2014ESP} and \cite{Tsay2014ICSE}, adding information about the geographical location of contributors and integrators. They conclude that PR acceptance rate is higher when both contributor and integrator are from the same country, with the exception of India, and that contributors from some countries (e.g., Switzerland and Japan) see their contributions more frequently accepted than contributors from other countries (e.g., China and Germany).

Although not directly related to PRs, Zhou and Mockus~\cite{Zhou2015} studied the acceptance behaviour of issue reports submitted to issue tracking systems of open source projects. They observed that low attention to submitted issues, as evidenced by a too-rapid (often negative) response reduces the chances of a newcomer becoming a long-term contributor.

\section{Methodology}

The main goal of our research is to study the longevity of PR-based contributions to large open source software projects. 
We focus on software development through 
\textsf{GitHub}, the largest and most active online hosting service for \textsf{git} projects. As of 2018-09-30, \textsf{GitHub} has hosted 96M+ repositories, 31M+ developers, and 200M+ PRs and about one third of these repositories and PRs were created in the last 12 months.\footnote{\url{https://octoverse.github.com}}

For this exploratory research, we selected three case studies of large open source \textsf{git}  projects on \textsf{GitHub}. 
These projects have been obtained by convenience sampling. This method is acceptable for getting preliminary research insights, and will be replaced in a later phase %by other sampling methods
to obtain a bigger corpus that covers a larger set of relevant projects.

The main criteria for our selected sample were that the projects should be representative of a typical PR-based software development process. To do so, the projects needed to be mature (i.e., have a time span of several years), have an active development history with a huge number of commits and contributors, and of course contain a very large number of PRs, in order to be able to derive statistically significant results from their analysis. In addition to this, we selected projects written in three different languages to ensure sufficient diversity. 
The three selected projects are \textsf{ansible}, \textsf{rails} and \textsf{kubernetes}. Some of their characteristics are shown in Table~\ref{RepoSummary}.

	\begin{table}[!htb]
		\centering
		\caption{Project repository characteristics on 24/10/2018}
		\label{RepoSummary}	
		\scalebox{0.9}{%
		\begin{tabular}{c|c|c|c|c|c}
%			\hline
			\textbf{repository} & \textbf{language} & \textbf{start year} & \textbf{\#contributors} & \textbf{\#commits} & \textbf{\#PR} \\
			\hline
			\textsf{ansible} & Python & 2012 & 3930 & 40k & 27k \\
			\textsf{rails} & Ruby & 2010 & 3683 & 70k & 22k \\
			\textsf{kubernetes} & Go & 2014 & 1861 & 71k & 42k \\
%			\hline
		\end{tabular}
	}
	\end{table}
	
Because we have observed problems of missing or inconsistent data when using GHTorrent, we decided to extract the PR data of the selected projects from \textsf{GitHub} repositories through the \textsf{GitHub} API directly. For each PR, the data contains information about the PR creation date, its status (accepted, rejected or open), its closing date (for accepted and rejected PRs), the \textsf{GitHub} ID of its author and its PR number. This PR number corresponds to a chronological ordering of issues opened in the repository, of which PRs are a subset.

\section{$RQ_1$: How are PR acceptance and rejection rates influenced by previous PRs?}

To answer RQ$_1$, we examined whether repeat contributions impact a contributor's PR acceptance rate.
To that effect, for each repository we analysed the PR acceptance rate in function of the number of submitted PRs by each contributor.

\begin{figure*}[!h]
	\centering
\begin{minipage}{.33\textwidth}
	\centering
	\includegraphics[width=\columnwidth]{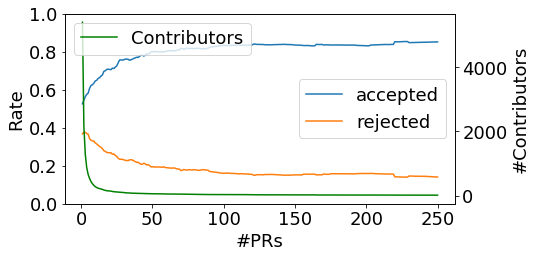}
	\caption*{\textsf{ansible}}
\end{minipage}%
\begin{minipage}{.33\textwidth}
	\centering
	\includegraphics[width=\columnwidth]{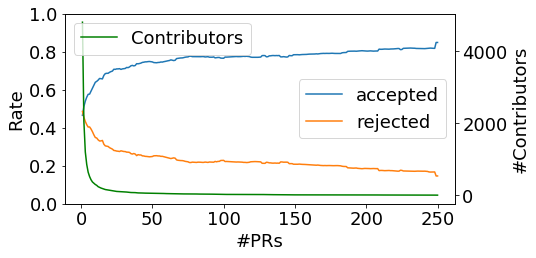}
	\caption*{\textsf{rails}}
\end{minipage}%
\begin{minipage}{.33\textwidth}
	\centering
	\includegraphics[width=\columnwidth]{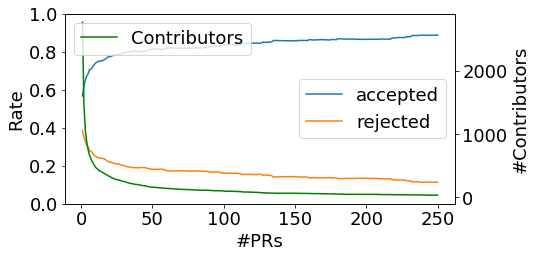}
	\caption*{\textsf{kubernetes}}
\end{minipage}%
	\caption{Acceptance rate of the first $x$ PRs of each contributor.}
	\label{AcceptanceExperience}
\end{figure*}

Figure~\ref{AcceptanceExperience} displays, for each positive integer threshold $x$ between 1 and 250, the PR acceptance rate (blue curve) and rejection rate (orange curve) considering the first $x$ PRs of each contributor only, thereby discarding contributors having less than $x$ PRs. The green curve shows the number of contributors having submitted at least $x$ PRs. Thresholds above 250 are excluded due to the low fraction of contributors having that many submissions: 0.33\% for  \textsf{ansible}, 0.15\% for \textsf{rails} and 1.23\% for \textsf{kubernetes}.

One can observe in all three examined project repositories that, as contributors submit more PRs, their acceptance rates increase significantly. 
Over the first 50 PRs, we observe a rise  from 54.2\% to 80.0\% for \textsf{ansible}, from 61.3\% to 81.4\% for  \textsf{rails}, and from 49.1\% to 74.3\% for \textsf{kubernetes}. Beyond the 50 first PRs, all three repositories saw continuous improvement in PR acceptance rates as contributors submitted more PRs to the project. 

\begin{figure*}[!h]
	\centering
	\begin{minipage}{.33\textwidth}
		\centering
		\includegraphics[width=\columnwidth]{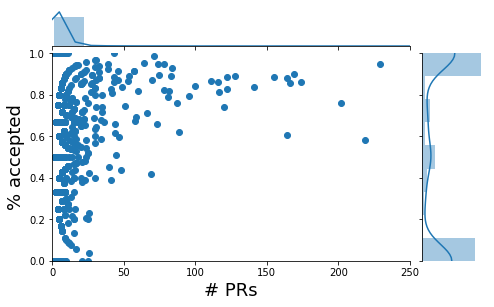}
		\caption*{\textsf{ansible}}
	\end{minipage}%
	\begin{minipage}{.33\textwidth}
		\centering
		\includegraphics[width=\columnwidth]{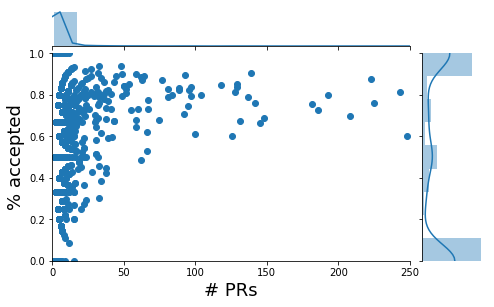}
		\caption*{\textsf{rails}}
	\end{minipage}%
	\begin{minipage}{.33\textwidth}
		\centering
		\includegraphics[width=\columnwidth]{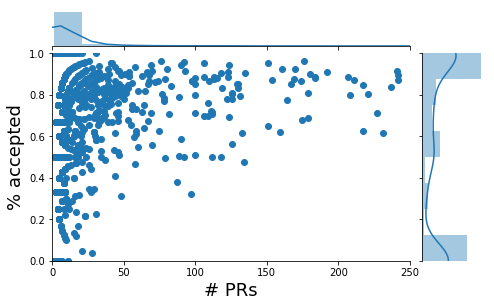}
		\caption*{\textsf{kubernetes}}
	\end{minipage}%
	\caption{Acceptance rate of all PRs by contributor.}
	\label{contribPRs}
\end{figure*}

These results agree with prior findings by Tsay et al~\cite{Tsay2014ICSE} and Gousios et al~\cite{Gousios2014ESP}, but are to be nuanced, given the rapid decrease in number of contributors as the threshold $x$ increases.
As a consequence, in Figure~\ref{contribPRs} we looked at the PR acceptance rate of all contributors, excluding the few that made over 250 contributions. We observe that, while contributors with a high number of PRs tend to have a consistently high PR acceptance rate, the behaviour for contributors with few PRs is quite unpredictable: they can have either low or high acceptance rates.
Therefore, although the number of previous PRs influences acceptance rate, this can only be verified starting from a certain threshold of PRs, below which no conclusion can be reached as to whether such an influence exists.

\section{$RQ_2$: To which extent does PR acceptance or rejection influence further contributions?}

While related work (e.g., \cite{Gousios2014ESP}) has studied the impact of PR acceptance rate on future PR decision time, $RQ_2$ focuses on the impact of PR acceptance rate on the likelihood of making further PRs.
To do so, we compared the probability to contribute again after either a rejected or an accepted PR. The results are presented in Table~\ref{recontributionTable}.
In all three considered projects, contributors are more likely to make subsequent PRs if their prior PRs were accepted.

\begin{table}[!htbp]
	\centering
	\caption{Likelihood to contribute again after PR acceptance or rejection.}
	\label{recontributionTable}	
	\begin{tabular}{l|c|c}
%		\hline
		\textbf{repository} & \textbf{after acceptance} & \textbf{after rejection} \\
		\hline
		\textsf{ansible} & 85.6\% & 73.0\% \\
		\textsf{rails} & 83.9\% & 69.2\% \\
		\textsf{kubernetes} & 95.5\% & 88.2\% \\
%		\hline
	\end{tabular}
\end{table}

We then used the statistical technique of survival analysis (a.k.a.
event history analysis)~\cite{Aalen2008}. 
Given a specific ``event of interest'' (in our case: acceptance or rejection of a PR),
survival analysis models the ``time to event'' data during a given observation period.
Survival functions model the survival rate of a given
dataset, i.e., the expected time duration until the event of
interest occurs. The models take into account the ``censoring'' of some observed
subjects, either because they enter or leave the
study during the observation period, or because the event of
interest was not observed for them during the observation
period. A common non-parametric statistic used to estimate
survival functions is the Kaplan-Meier estimator~\cite{KaplanMeier2012}.

\begin{figure*}[!h]
	\centering
\begin{minipage}{.33\textwidth}
	\centering
	\includegraphics[width=\columnwidth]{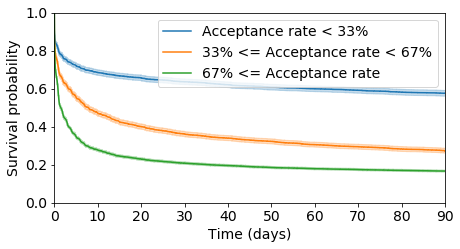}
	\caption*{\textsf{ansible}}
\end{minipage}%
\begin{minipage}{.33\textwidth}
	\centering
	\includegraphics[width=\columnwidth]{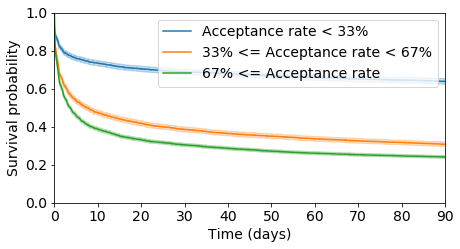}
	\caption*{\textsf{rails}}
\end{minipage}%
\begin{minipage}{.33\textwidth}
	\centering
	\includegraphics[width=\columnwidth]{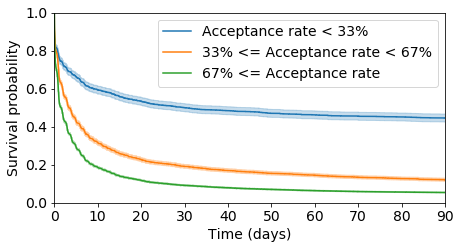}
	\caption*{\textsf{kubernetes}}
\end{minipage}%
	\caption{Survival curves for the probability to submit a next PR, grouped by acceptance rate classes.}
	\label{survivalDelayAcceptance}
\end{figure*}

We performed an analysis of the survival probability to submit a new PR in function of the time elapsed since the latest submission (at that time) of a PR by the same contributor.
In order to assess if the PR acceptance rate influences the delay for a contributor to submit new PRs, we considered three acceptance rate classes: $[0,33\%[,[33\%,67\%[ \text{ and } [67\%,100\%]$. 
The survival curves are shown in Figure~\ref{survivalDelayAcceptance}. 

We observe that the survival probability is higher for classes of higher acceptance rate, regardless of the considered projects. For instance, after ten days, the probability to submit a new PR is 72.2\% in \textsf{Ansible} if the acceptance rate is over 67\%, while this probability drops to 52.9\% if the acceptance rate is between 33\% and 67\%, and even to 31.5\% if the acceptance rate is below 33\%. Similar patterns can be observed for the two other projects. 

We carried out pairwise log-rank tests to compare whether statistically significant differences could be found between the survival curves. The differences were statistically confirmed at $\alpha = 0.01$ (after a Bonferroni correction \cite{Haynes2013}), i.e., the null hypotheses, assuming that the survival curves for different acceptance rate classes are the same, were rejected.

We performed a proportional hazards regression based on Cox regression to determine to which extent the acceptance rate impacts the probability of further contribution~\cite{Cox1972}. The Cox regression is a method for investigating the effect of several variables upon a specified event's hazard rate. For this analysis, we included the following factors: the acceptance rate of all prior PRs by the same contributor; the number of prior PRs made by this contributor, and the time elapsed since the contributor's first PR (the contributor's \textit{age}).

\begin{table}[!htbp]
	\centering
	\caption{Influence of acceptance rate, number of PRs, and contributor age on the time required to submit a new PR.}
	\label{coxTable}	
%		\scalebox{0.85}{%
	\begin{tabular}{l|c|c|c|c}
%		\hline
		  & \multicolumn{3}{c|}{\textbf{regression coefficients for}} & \\ 
%		\cline{2-4} 
		\textbf{repository} & \textbf{acceptance} & \textbf{\#prior} & \textbf{contributor} & \textbf{concordance} \\
		\textbf{} & \textbf{rate} & \textbf{PRs} & \textbf{age} & \textbf{} \\
		\hline
		\textsf{ansible} & 0.5481 & 0.0038 & -0.0009 & 0.689 \\
		\textsf{rails} & 0.4630 & 0.0047 & -0.0015 & 0.728 \\
		\textsf{kubernetes} & 0.7455 & 0.0031 & -0.0016 & 0.637 \\
%		\hline
	\end{tabular}%
%	}
\end{table}

Table~\ref{coxTable} summarizes the results we obtained. The concordance (fourth column) provides the goodness of fit of the model. It is comprised between 0 (perfect anti-concordance) and 1 (perfect concordance). The table also reports the regression coefficients for the three considered factors. These coefficients measure the magnitude of the impact of the aforementioned factors on the probability to submit a new PR. All these coefficients are statistically significant ($p<0.01$ after Bonferroni correction). Their values signify that an increase of one increment (10\% in acceptance rate, 1 prior PR or 1 day since the contributor's first PR) multiplies the probability to submit a PR by a factor $e^{\textit{coefficient}}$.  
For instance, in the case of Kubernetes: an increase of 10\% in PR acceptance rate modifies the probability to submit a new PR by a factor 1.0774, each prior PR by 1.0032 and each day since the contributor's first PR by 0.9984.

\section{$RQ_3$: To which extent do PRs left open influence further contributions?}

\begin{figure*}[!h]
	\centering
\begin{minipage}{.33\textwidth}
	\centering
	\includegraphics[width=\columnwidth]{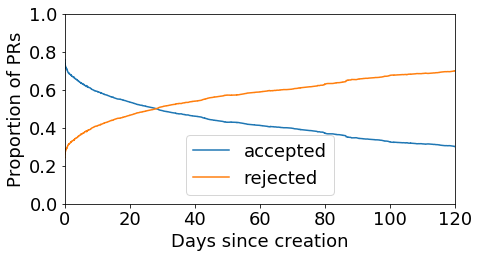}
	\caption*{\textsf{ansible}}
\end{minipage}%
\begin{minipage}{.33\textwidth}
	\centering
	\includegraphics[width=\columnwidth]{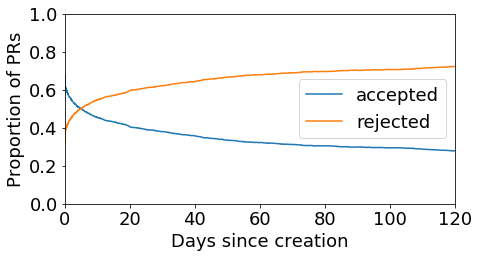}
	\caption*{\textsf{rails}}
\end{minipage}%
\begin{minipage}{.33\textwidth}
	\centering
	\includegraphics[width=\columnwidth]{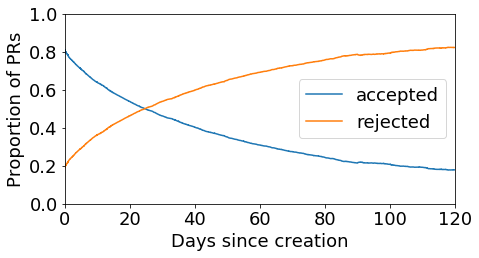}
	\caption*{\textsf{kubernetes}}
\end{minipage}%
	\caption{Proportion of PRs that were ultimately accepted in function of their age}
	\label{RQ3}
\end{figure*}

To provide insight into $RQ_3$, we looked at the proportion of PRs that were ultimately accepted or rejected given the time it took to decide (the PR's \textit{age}). We excluded PRs that were left open, since no decision has been reached for those.
This is plotted in Figure~\ref{RQ3}. We notice that, the longer a PR remains open, the higher the probability that it will be rejected. After a threshold $x$, PRs have a higher probability to be rejected than accepted. The threshold is 28 days for \textsf{ansible}, 5 days for \textsf{rails} and 25 days for \textsf{kubernetes}. Presuming that contributors are aware of this phenomenon, we expect that they implicitly consider PRs left open for a too long duration as being \emph{tacitly} rejected, producing effects similar to those identified in the previous RQ.
This preliminary result needs to be confirmed with further analyses.

\section{Threats to validity}
A threat to the validity of this paper is the fact that we only selected three projects in this exploratory phase, so the preliminary findings might not generalise to bigger sets of projects. Choosing only large, popular and mature projects are also a source of bias, as Rahman and Roy \cite{Rahman2014msr} found that such factors affect PR acceptance rates.

Another threat is that the PR status returned by the \textsf{GitHub} API does not necessarily correspond to the actual fate of the PR in some repositories. One such case is \textsf{homebrew-core}, where the policy of the repository is to close most PRs without merging them, but to integrate those they deem appropriate through another mechanism, such as the integrators committing the changes themselves.\footnote{https://docs.brew.sh/How-To-Open-a-Homebrew-Pull-Request}
If analyses were to be applied to this repository, the rate of acceptance would be artificially low due to that specific PR handling policy.
Another example is that of \textsf{angular}, wherein PRs marked with specific tags (``PR action: merge'' and ``PR target: *'' where * represents one or more branch branches to merge the PR into) will have their relevant code automatically integrated into the repository through commits. Those PRs will appear to be rejected on \textsf{GitHub}, even though they aren't.
It would be possible to recover the actual PR status based on those tags, which is not the case for \textsf{homebrew-core}.

Yet another threat is tied to the way we have identified contributors.
It has been reported that a single individual may use multiple identities in different capacities or at different times on software repositories~\cite{GoeminneComparison2012,Kouters2012Who,Wiese2016ICSME}. 
More specifically, it may be the case that the same author owns multiple \textsf{GitHub} accounts, or even that multiple authors contribute under the same \textsf{GitHub} account.
In that case, we may have erroneously attributed the PRs of a contributor to his identities.
This could have affected our findings. Therefore, as future work, we aim to empirically study the impact of such incorrect contributor identification.

\newpage
\section{Conclusion}

The collaborative development of open-source software through a pull-based contribution process involves subtle social interactions that can influence the frequency and likelihood of contribution to a repository, or even its ability to retain contributors.
Recent qualitative results have highlighted that contributors do not appreciate the rejection of their PRs, and that they find poor responsiveness from integrators frustrating. Integrators, on the other hand, are wary of alienating contributors in their handling of PRs. 

In this paper, we provide preliminary quantitative results showing that a contributor's PRs are more more likely to be accepted when he has submitted more PRs previously.
We also reveal the impact of PR decisions on the willingness of contributors to contribute anew. Indeed, fewer contributors submit a new PR after the previous one was rejected than when the previous one was accepted.
This highlights the importance for project integrators to avoid aleniating contributors, lest they lose their contributions.

\section*{Acknowledgements}
This research was supported by the FRQ-FNRS collaborative research project R.60.04.18.F \emph{SECOHealth}, the Excellence of Science project 30446992 \emph{SECO-ASSIST} financed by FWO-Vlaanderen and F.R.S.-FNRS, and F.R.S.-FNRS Grant T.0017.18.

\newpage
\balance
\providecommand{\noopsort}[1]{}

\end{document}